\documentclass[aps,pra,twocolumn,groupedaddress]{revtex4-1}
\usepackage{xcolor,graphicx,ulem}
\usepackage{amsmath, amssymb}
\usepackage[colorlinks=true,urlcolor=blue,citecolor=blue,linkcolor=magenta]{hyperref}

\usepackage{tikz} 
\usetikzlibrary{fit,matrix} 
\usepackage{braket} 
\usepackage{pgfplots}
\usepackage[miktex]{gnuplottex}
\usepackage[subpreambles=true]{standalone}

\begin{document}

\title{Verifying bound entanglement of dephased Werner states}

\author{P. Thomas}
\author{M. Bohmann}\email{martin.bohmann@uni-rostock.de}
\author{W. Vogel}
\affiliation{Institut f\"ur Physik, Universit\"at Rostock, Albert-Einstein-Str. 23, D-18051 Rostock, Germany}

\begin{abstract}

	The verification of quantum entanglement under the influence of realistic noise and decoherence is crucial for the development of quantum technologies.
	Unfortunately, a full entanglement characterization is generally not possible with most entanglement criteria such as entanglement witnesses or the partial transposition criterion.
	In particular, so called bound entanglement cannot be certified via the partial transposition criterion.
	Here we present the full entanglement verification of dephased qubit and qutrit Werner states via entanglement quasiprobabilities.
	Remarkably, we are able to reveal bound entanglement for noisy-mixed states in the qutrit case.
	This example demonstrates the strength of the entanglement quasiprobabilities for verifying the full entanglement of quantum states suffering from noise. 
	
\end{abstract}

\date{\today}

\maketitle

\section{Introduction}
	
	Quantum entanglement is an important resource for quantum information \cite{Nielsen2000}.
	However this resource will be affected by the environment in real applications.
	In particular noise has a negative influence on such quantum systems \cite{Zurek2003,Schlosshauer2005}.
	Thus some questions arise regarding the usefulness of an entangled state for quantum tasks after such interactions and the resistance of entanglement in the presence of noise.
	
	Various entanglement criteria and strategies for entanglement verification have been developed \cite{Horodecki2009}.
	However, most entanglement criteria have the disadvantage that they do not provide a necessary and sufficient condition for entanglement.
	In other words, they only certify entanglement if the specific condition is violated, while the non-violation does not allow a conclusion concerning separability.
	Among such conditions are the so called entanglement witnesses \cite{Horodecki1996}.
	Entanglement witnesses certify entanglement if the expectation value of the witness operator is negative.
	Therefore, the outcome of the entanglement test depends on the considered quantum state and the choice of the specific witness operator. 
	Note that the construction and optimization of witnesses is an important and ongoing task, see, e.g., \cite{Lewenstein2000,Sperling2009,Shahandeh2017}.
	
	Another well established and widely used entanglement criterion is the partial transposition criterion \cite{Peres1996}.
	The partial transpose of a separable state is positive, and thus a negative partial transposition (NPT) reveals entanglement.
	The positivity of the partial transposition is necessary and sufficient for separability in systems of dimensions $2\otimes 2$, $2\otimes 3$, and Gaussian bipartite states \cite{Simon2000} only, whereas for systems of dimensionality of $3\otimes 3$ and higher it is only a necessary one \cite{Horodecki1996}.
	Consequently, there exist entangled states which exhibit a positive partial transposition.
	Such states are also referred to as bound entangled states \cite{Horodecki1998} and they are intensively studied in the context of entanglement distillation, see, e.g., \cite{Horodecki1998b,Duer2000,Vollbrecht2002}.
	Entanglement of states with positive partial transposition, however, may be certified via other conditions such as the realignment criterion \cite{Sharma2016} or via entanglement witnesses \cite{Hyllus2004}.

	One way of doubtlessly identifying all entanglement of a quantum state can be achieved via the concept of entanglement quasiprobabilities \cite{Sperling2009QP}.
	The entanglement quasiprobability representation is an expansion of the quantum state in a separable-state-diagonal form which unambiguously certifies separability by its non-negativity. 
	Accordingly, we certify entanglement via negativities in the entanglement quasiprobability distribution.
	We would like to emphasize the conceptual similarity between the entanglement quasiprobabilities and the Glauber-Sudarshan $P$ representation \cite{Glauber,Sudarshan} for nonclassicality.
	Note that entanglement quasiprobabilities have been applied to two-mode squeezed-vacuum states \cite{Sperling2012} and generalized N00N states \cite{Bohmann2017}.	

	As an example state for our studies we consider the family of Werner states \cite{Werner1989}.
	These are a set of one parameter states which are invariant under local unitary transformations.
	Werner states are of particular importance due to the fact that any NPT state can be mapped onto an NPT Werner state with local operations and classical communication \cite{Horodecki1999}.
	Since they cover separable, mixed, and pure entangled states, the Werner states have been studied for experimental entanglement detection \cite{Barbieri2003}.
	Werner states can also be used for quantum tasks like entanglement teleportation \cite{Lee2000}. 
	They have also been analyzed with respect to the loss of entanglement under dimensional reduction \cite{Moelmer2004}. 
	In order to generate such states parametric down conversion \cite{Zhang2002,Cinelli2004} or a collective decay of atoms \cite{Agarwal2006} can be used.
	Note that we also studied entanglement quasiprobabilities of noisy N00N states in Ref. \cite{Bohmann2017}. 
	In the present case, however, we directly demonstrate the verification of bound entanglement by entanglement quasiprobabilities. 
	This serves as a proof of principle and demonstrates the power of this method.
	
	In this paper, we investigate the verification of entanglement of dephased Werner states.
	Therefore, we calculate the Werner states undergoing a dephasing channel.
	Firstly, we apply the NPT criterion to the dephased Werner states which serves as the reference for bound entanglement.
	Secondly, we formulate the separability eigenvalue problem for the dephased Werner states and give solutions for these coupled eigenvalue equations for the qubit and qutrit case.
	Furthermore, we calculate optimal entanglement quasiprobabilities based on these solutions.
	The results for the partial transposition criterion and the negativities of the obtained entanglement quasiprobability distributions are compared.
	In particular, we identify parameter regions of bound entanglement for the qutrit Werner states.
	
	The paper is organized as follows.
	In Sec. \ref{ch:Werner} we consider the general dephased qudit Werner states and investigate their entanglement with the partial transposition criterion for the qubit and qutrit case.
	The verification of entanglement with entanglement quasiprobabilities for the qubit and qutrit Werner states with general dephasing and especially Gaussian dephasing will be demonstrated in Sec. \ref{ch:QP}.
	We summarize and conclude in Sec. \ref{ch:Conclusion}.
	
\section{Werner states in dephasing environments}\label{ch:Werner}

	In this section we firstly consider the qudit Werner state \cite{Werner1989}.
	Secondly, we calculate the dephased qudit Werner states for a general single-mode dephasing channel.
	Finally, we apply the partial transposition criterion to certify the entanglement of the dephased qubit and qutrit Werner states.
	
\subsection{Werner states}\label{ch:State}

	The family of Werner states, introduced in Ref \cite{Werner1989}, includes separable, mixed entangled and pure entangled states.
	In this paper we use the following representation of the qudit Werner states:
	\begin{equation}\label{eq:State}
		\hat\rho_W=\mathcal{N}(\alpha)\left(\hat 1-\alpha \hat S\right)
	\end{equation}
	with $\alpha\in[-1,1]$, the swap operator $\hat S$,
	\begin{equation}\label{eq:Swap}
		\hat S=\sum_{i,j}\ket{i,j}\bra{j,i},
	\end{equation}	
	and the normalization factor $\mathcal{N}(\alpha)=\left(d^2-\alpha d\right)^{-1}$ with the dimension $d$.
	As mentioned before, Werner states are invariant under local unitary transformations
	\begin{equation}
		\hat\rho_W=(\hat U\otimes \hat U)\hat\rho_W(\hat U\otimes \hat U)^\dagger,
	\end{equation}
	which can be verified for the state given in Eq. \eqref{eq:State}.
	Due to their construction, the Werner states are entangled if \cite{Werner1989}
	\begin{equation}
	\alpha>\frac{1}{d}.
	\end{equation}
	As we are only interested in the entanglement property of the Werner states, we will only consider states with $\alpha>0$ in the following calculations.

\subsection{Dephased Werner states}\label{ch:Dephasing}

	A two-mode dephasing channel can be described via the transformation
	\begin{align}\label{eq:DephasingMap}
		\hat\rho\mapsto\Lambda_{\rm deph}(\hat\rho)=&\int_{0}^{2\pi} d\varphi_a\int_{0}^{2\pi} d\varphi_b\, p(\varphi_a,\varphi_b)
		\\&\times\nonumber
		[e^{i\varphi_a\hat n}\otimes e^{i\varphi_b\hat n}]
		\hat\rho
		[e^{-i\varphi_a\hat n}\otimes e^{-i\varphi_b\hat n}],
	\end{align}
	with the phase distribution $p(\varphi_a,\varphi_b)$ and $\hat n$ is the photon number operator in the corresponding mode.
	Note that $p(\varphi_a,\varphi_b)$ has the properties of a classical probability distribution and Eq. \eqref{eq:DephasingMap} describes a trace-preserving quantum channel.
	Applying Eq. \eqref{eq:DephasingMap} to the Werner states yields the dephased Werner states $\hat\rho_{\rm{W,deph}}$,
	\begin{equation}\label{eq:WernerDephasing}
		\hat\rho_{\mathrm{W,deph}}=\mathcal{N}(\alpha)\left(\hat 1-\alpha \sum_{m,n}\lambda_{mn}\ket{m,n}\bra{n,m}\right),
	\end{equation}
	with 
	\begin{equation}\label{eq:Lambda}
	\lambda_{m,n}=\int_{0}^{2\pi} d\varphi_a\int_{0}^{2\pi} d\varphi_b\, p(\varphi_a,\varphi_b) e^{i(\varphi_a-\varphi_b) (m-n)}.
	\end{equation}
	From the Hermiticity follows that $\lambda_{mn}$=$\lambda_{nm}^*$.
	Furthermore, the dephasing yields a reduction of the off-diagonal elements $\lambda_{mn}~(m\not=n)$, while the diagonal elements are preserved, $\lambda_{mm}=1$.
	Note, however, that the Werner states and the dephased Werner states have the same algebraic structure [see, Eqs. \eqref{eq:WernerDephasing} and \eqref{eq:State}].

	We may, specifically, consider the case of Gaussian dephasing in the second mode (see also Refs. \cite{Sperling2012,Bohmann2015,Bohmann2017}).
	In this case, the phase distribution reads as $p(\varphi_a,\varphi_b)=\delta_A(\varphi_a)p_B(\varphi_b)$, where $\delta_A$ denotes the Dirac-delta distribution and the phase distribution $p_B(\varphi_b)$ can be described by a wrapped Gaussian phase distribution,
	\begin{align}\label{eq:Gauss}
		p_B(\varphi_b)=\sum\limits_{k\in \mathbb Z}\frac{1}{\sqrt{2\pi\delta^2}}\exp\left[-\frac{(\varphi_b+2k\pi)^2}{2\delta^2}\right],
	\end{align}
	where $\delta$ is the width of the initial Gaussian distribution.
	For this specific distribution we obtain the coefficients 
	\begin{equation}\label{eq:CoeffGauss}
		\lambda_{mn}=e^{-\delta^2(m-n)^2/2}.
	\end{equation}
	Note that in the case of $\delta\to\infty$ the Gaussian distribution becomes a uniform distribution which leads to the loss of the offdiagonal terms, $\lambda_{mn}=0$ for $m\neq n$.
	
	So far we obtained the qudit Werner states undergoing dephasing.
	For the specific case of Gaussian dephasing we calculated the dephasing coefficients.
	In the following, we focus on the entanglement of qubit and qutrit Werner states under such dephasing conditions.

\subsection{Partial transposition of dephased Werner states}\label{ch:PT}
	
	Let us now study the entanglement of dephased qubit and qutrit Werner states with the partial transposition criterion \cite{Peres1996}.
	The partial transposition in mode B for a bipartite state $\hat\rho=\sum_{ijkl}p_{ijkl}|i\rangle\langle j |\otimes |k\rangle\langle l |$ reads as
	\begin{align}\label{eq:PT}
		\begin{split}
			\hat\rho^{\rm T_{\rm B}}=\sum_{ijkl}p_{ijkl}|i\rangle\langle j |\otimes |l\rangle\langle k |.
		\end{split}
	\end{align}
	The partial transposition of any separable state has positive eigenvalues.
	Thus positivity of this map is a necessary condition for separability.
	For $2\times2$ and $2\times3$ systems the positivity is also a sufficient criterion \cite{Horodecki1996}.
	
	Applying the partial transposition on mode B of the general dephased Werner states in Eq. \eqref{eq:WernerDephasing} results in
	\begin{equation}\label{eq:WernerDephPT}
		\hat\rho_{W,deph}^{T_B}=\mathcal{N}(\alpha)\left(\hat1-\alpha\sum_{m,n}\lambda_{mn}\ket{m,m}\bra{n,n}\right).
	\end{equation}
	The resulting partially transposed density matrix $\hat\rho_{W,deph}^{T_B}$ is negative if we find that it has at least one negative eigenvalue.
	Therefore, we calculate its eigenvalues for the qubit ($d=2$) and qutrit ($d=3$) case.
	In the qubit case this yields
	\begin{align}\label{eq:PTEigenvalD2}
		\begin{split}
			\mathcal{N}(\alpha)\left(1,1,1-\alpha+\alpha|\lambda_{01}|,1-\alpha-\alpha|\lambda_{01}|\right).
		\end{split}
	\end{align}
	In the qutrit case we get
	\begin{align}\label{eq:PTEigenval}
		\begin{split}
			\mathcal{N}(\alpha)\bigg(1,1,1,1,1,1,1-\alpha+\alpha|\lambda_{02}|,\\
			\frac{1}{2}\left(2-2\alpha-\alpha|\lambda_{02}|-\alpha\sqrt{8|\lambda_{01}|^2+|\lambda_{02}|^2}\right),\\
			\left.\frac{1}{2}\left(2-2\alpha-\alpha|\lambda_{02}|+\alpha\sqrt{8|\lambda_{01}|^2+|\lambda_{02}|^2}\right)\right).
		\end{split}
	\end{align}
	In order to verify entanglement by the NPT criterion we need to consider for which conditions negative eigenvalues can be found.
	Note that this depends on the dephasing properties given by the coefficients $\lambda_{m n}$ [see Eq. \eqref{eq:Lambda}] and the state parameter $\alpha$.
	From Eq. \eqref{eq:PTEigenvalD2} we find that the qubit state has a negative partial transposition if
	\begin{equation}\label{eq:ANPTD2}
		\alpha^{\rm PT}_{d=2}=\frac{1}{1+|\lambda_{01}|}.
	\end{equation}
	Here, the superscript ${\rm PT}$ indicates that the bound is obtained via the PT criterion.
	For the case of Gaussian dephasing [cf. Eqs. \eqref{eq:Gauss} and \eqref{eq:CoeffGauss}] Eq. \eqref{eq:ANPTD2} reads as
	\begin{equation}\label{eq:ANPTD2Gauss}
		\alpha^{\rm PT}_{d=2}=\frac{1}{1+e^{-\delta^2/2}}.
	\end{equation}
	Similarly, we obtain the condition for the NPT of the qutrit state from Eq. \eqref{eq:PTEigenval}:
	\begin{equation}\label{eq:ANPT}
		\alpha^{\rm PT}_{d=3}=\frac{2}{2+|\lambda_{02}|+\sqrt{8|\lambda_{01}|^2+|\lambda_{02}|^2}}.
	\end{equation}
	Again for Gaussian dephasing we obtain the specific condition
	\begin{equation}\label{eq:ANPTGauss}
		\alpha^{\rm PT}_{d=3}=\frac{2}{2+e^{-2\delta^2}+\sqrt{8e^{-\delta^2}+e^{-4\delta^2}}}.
	\end{equation}
	If $\alpha$ exceeds the corresponding lower limit the state exhibits NPT entanglement.
	Since we considered general dephasing any phase distribution of interest can be used to calculate the dephasing coefficients.
	
	To this end, let us stress that all entangled Werner states show NPT entanglement \cite{Horodecki1999}.
	Therefore, the NPT condition is a necessary and sufficient entanglement condition for Werner states.
	However, this is not the case for dephased Werner states as they are a mixture of Werner states [see Eq. \eqref{eq:DephasingMap}].

\section{Verifying entanglement with entanglement quasiprobabilities}\label{ch:QP}

	In this section we derive the entanglement quasiprobabilities for the dephased Werner state.
	Negativities in the entanglement quasiprobabilities are a necessary and sufficient condition for entanglement of the considered state.
	In order to obtain the entanglement quasiprobabilities of the family of dephased Werner states, we will solve the underlying separability eigenvalue problem.
	With the solutions of the separability eigenvalue problem we calculate the entanglement quasiprobability distributions for the qubit and qutrit case with general and especially Gaussian dephasing.
	Finally, we compare the entanglement verification by entanglement quasiprobabilities with that by partial transposition.

\subsection{The separability eigenvalue problem and entanglement quasiprobabilities}

	For a given bipartite Hermitian operator $\hat L$ the SEP can be given as \cite{Sperling2009,Sperling2013}
	\begin{equation}\label{eq:SEP2}
		\hat L\ket{a,b}=g\ket{a,b}+\ket{\chi}
	\end{equation}
	with the separability eigenvalue $g$, the corresponding separability eigenvector $\ket{a,b}$, and a biorthogonal perturbation $\ket{\chi}\sim |a^\perp,b^\perp\rangle$ with $\langle a|a^\perp\rangle=\langle b|b^\perp\rangle=0$.
	The SEP in the form of Eq.~\eqref{eq:SEP2} has the form of a disturbed eigenvalue problem.
	Note that the SEP can also be formulated as a coupled system of eigenvalue equations (see \cite{Sperling2009,Sperling2013}).
	Solving Eq. \eqref{eq:SEP2} gives a set of eigenvalues and -vectors
	\begin{equation}\label{eq:SEPSolutions}
		\{g_i,\ket{a_i,b_i}\}_i.
	\end{equation}
	The maximal separability eigenvalue $g_{\mathrm{max}}$ allows one to construct an optimal entanglement witness $\hat W=g_{\mathrm{max}}\hat 1-\hat L$ (see \cite{Sperling2009}).
	Note that this method has been extended to multipartite scenarios \cite{Sperling2013} and has been successfully applied to experimental data \cite{Gutierrez2014,Gerke2015,Gerke2016}.
	
	We will now briefly show how the solutions of Eq. \eqref{eq:SEP2} can be used in order to obtain the entanglement quasiprobabilities of a bipartite quantum state \cite{Sperling2009QP}.
	Negativities in the entanglement quasiprobabilities then unambiguously reveal the entangled character of the state, while separable states exhibit nonnegative entanglement quasiprobabilities.
	For the derivation and a detailed analysis of the concept of entanglement quasiprobabilities we refer to Ref. \cite{Sperling2009QP}.
	
	Let us now consider the bipartite quantum state $\hat \rho$ and the solutions to its SEP ($\hat L=\hat \rho$) given by Eq. \eqref{eq:SEPSolutions}.
	With the separability eigenvectors one can calculate the matrix $\textbf{G}=(|\braket{a_i,b_i|a_j,b_j}|^2)_{ij}$.
	Together with the vector of separability eigenvalues $\vec{g}=(g_i)_i$ one obtains the following equation
	\begin{equation}\label{eq:QPS}
		\textbf{G}\vec{p}=\vec{g}
	\end{equation}
	where $\vec{p}$ has the form of an entanglement quasiprobability distribution.
	However the solution $\vec{p}$ of Eq. \eqref{eq:QPS} may be ambiguous, due to the zero eigenvalues of $\bf{G}$, and therefore, it has to be optimized.
	For this step one needs the kernel of $\bf{G}$ to calculate an optimized entanglement quasiprobability distribution $P_{\mathrm{Ent}}$ (see \cite{Sperling2009QP}),
	\begin{equation}\label{eq:OPT}
		P_{\mathrm{Ent}}\cong(P_{\mathrm{Ent}}(a_l,b_l))_l=\vec{p}-\sum_k\frac{\vec{p}_{0_k}^T\vec{p}}{\vec{p}_{0_k}^T\vec{p}_{0_k}}\vec{p}_{0_k}
	\end{equation}
	with $\mathrm{ker}(\textbf{G})=\{\vec{p}_{0_k}\}_k$.
	Finally, one has an optimal entanglement quasiprobability distribution the negativities of which are a necessary and sufficient signature of its entanglement.
	In the following sections we apply the method of entanglement quasiprobabilities to the general dephased Werner states.

\subsection{SEP of dephased Werner states}\label{ch:SEP}
	
	After introducing the concepts, we will now solve the SEP for the family of dephased Werner states which eventually allows us to construct their entanglement quasiprobabilities.
	For detailed information about the solutions and the choice of separability eigenvectors, we refer to the Appendix \ref{app:Appendix}.
	Formulating the SEP \eqref{eq:SEP2} for the general dephased Werner states of the form Eq. \eqref{eq:WernerDephasing} yields
	\begin{equation}\label{eq:WernerSEP}
		\hat\rho_{\mathrm{W,deph}}\ket{a,b}=\mathcal{N}(\alpha)\left(\ket{a,b}-\alpha\sum_{i,j}\lambda_{ij}\ket{i,j}\braket{j,i|a,b}\right).
	\end{equation}
	For simplicity we set $\lambda_{ij}\in\mathbb{R}$.
	In this way, we obtain immediately the trivial separability eigenvectors and the corresponding separability eigenvalues
	\begin{equation}\label{eq:SEPTriv}
		\ket{a,b}=\ket{i,j} 
		\quad
		\text{and}
		\quad
		g_{ij}=\mathcal{N}(\alpha)\left(1-\delta_{ij}\alpha\right),
	\end{equation}
	respectively.
	A meaningful choice for the nontrivial separability eigenvectors is $\ket{\mathfrak{s}_{jk}^l,\mathfrak{s}_{jk}^l}$, $\ket{\mathfrak{s}_{jk}^l,\mathfrak{s}_{jk}^{l+2}}$, $\ket{\mathfrak{s}_{jk}^l,m}$, and $\ket{m,\mathfrak{s}_{jk}^l}$ with $m\neq j,k$ and
	\begin{equation}\label{eq:SEVStructure}
		\ket{\mathfrak{s}_{jk}^l}=\frac{1}{\sqrt{2}}(\ket{j}+i^l\ket{k}).
	\end{equation}
	Note that vectors given by Eq. \eqref{eq:SEVStructure} are eigenvectors of the generalized Gell-Mann matrices and span together with the vectors $\ket{i}$ for $i=0,...,d-1$ the hyperplane of quantum states of the Hilbert space $\mathcal{H}=\mathbb{C}^d$.
	Then, the separability eigenvectors have the following separability eigenvalues
	\begin{align}\label{eq:SEPNonTriv}
		\begin{aligned}
			\ket{\mathfrak{s}_{jk}^l,\mathfrak{s}_{jk}^l}&:&g_{jk}^\parallel&=\mathcal{N}(\alpha)\left(1-\frac{\alpha}{2}\left(1+|\lambda_{jk}|\right)\right),\\
			\ket{\mathfrak{s}_{jk}^l,\mathfrak{s}_{jk}^{l+2}}&:&g_{jk}^\perp&=\mathcal{N}(\alpha)\left(1-\frac{\alpha}{2}\left(1-|\lambda_{jk}|\right)\right),\\
			\ket{\mathfrak{s}_{jk}^l,m},\ket{m,\mathfrak{s}_{jk}^l}&:&g_{jkm}&=\mathcal{N}(\alpha).
		\end{aligned}
	\end{align}
	The dephasing actually leads to slightly less degenerated separability eigenvectors compared to the Werner states without dephasing.
	These solutions can also be applied for any dimension $d$, but for $d>3$ additional separability eigenvectors may be necessary to calculate the entanglement quasiprobability distribution.
	In order to obtain the entanglement quasiprobability distribution for dephased Werner states with $d>3$, one can follow the procedure presented in this paper.
	The solutions of the SEP of the Werner states for arbitrary $d$ values without dephasing are based on the eigenvectors of the generalized Gell-Mann matrices.
	On this basis we calculate the corresponding separability eigenvalues of the dephased Werner states by inserting these separability eigenvectors into the SEP.
	Further information concerning the choice of separability eigenvectors in the qubit and qutrit case are given in the Appendix \ref{app:Appendix}, which can be generalized to $d>3$ in a straightforward manner.

\subsection{Entanglement quasiprobabilities for dephased Werner states}\label{ch:QPQutrit}

	In this subsection we explicitly calculate the entanglement quasiprobabilities of the qubit and qutrit dephased Werner states.
	Especially in the case of a Gaussian phase distribution, we compare the entanglement verified via NPT and negativities in the entanglement quasiprobabilities.
	
	\begin{figure}[h]
		\centering
		\includegraphics[width=\columnwidth]{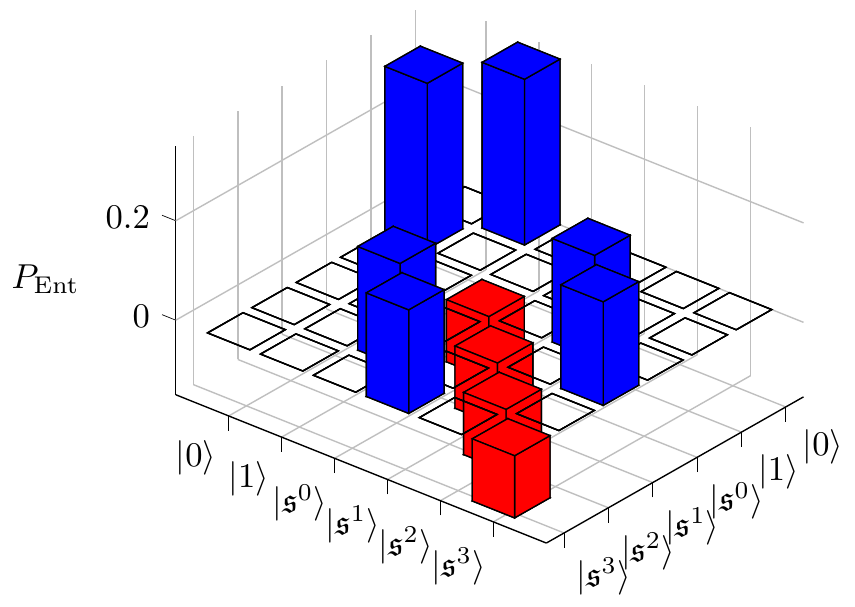}
			\caption{
				The entanglement quasiprobability distribution for the qubit Werner state for $\alpha=0.8$, without dephasing.
				Each bar represents the value of the entanglement quasiprobability distribution $P_{\mathrm{Ent}}(a,b)$ for the vector $\ket{a,b}$.
				The negativities indicate the entanglement of the given state.
			}
		\label{fig:WDepD2}
	\end{figure}
	
	In order to obtain the entanglement quasiprobabilities we use the separability eigenvectors and separability eigenvalues in Eq. \eqref{eq:SEPNonTriv}, which allows us to formulate the linear system of Eqs. \eqref{eq:QPS}.
	Solving this set of equations and optimizing $\vec{p}$ according to Eq. \eqref{eq:OPT} leads to the following optimized entanglement quasiprobability distribution for the dephased qubit Werner state
	\begin{align}
		P_{\mathrm{Ent}}\cong\left(p_{01},p_{01},p_{01}^\parallel,p_{01}^\parallel,p_{01}^\parallel,p_{01}^\parallel,p_{01}^\perp,p_{01}^\perp,p_{01}^\perp,p_{01}^\perp\right)
	\end{align}
	with the entries
	\begin{align}\label{eq:QPQubit}
		\begin{aligned}
			p_{01}&=\mathcal{N}(\alpha)\alpha,\\
			p_{01}^\parallel&=\frac{\mathcal{N}(\alpha)}{2}\left[1-\alpha\left(1+|\lambda_{01}|\right)\right],\\
			p_{01}^\perp&=\frac{\mathcal{N}(\alpha)}{2}\left[1-\alpha\left(1-|\lambda_{01}|\right)\right].
		\end{aligned}
	\end{align}
	Here, the corresponding separability eigenvectors for entries without superscript refer to the trivial separability eigenvectors in Eq. \eqref{eq:SEPTriv}, and the superscripts $\parallel$ and $\perp$ refer to the separability eigenvectors in Eqs. \eqref{eq:SEPNonTriv}.
	In Fig. \ref{fig:WDepD2} the representation of the qubit Werner state with product states and an entanglement quasiprobability distribution is shown for $\alpha=0.8$ and no dephasing.
	The smallest value of the entanglement quasiprobability distribution in the qubit case is $p_{01}^\parallel$
	which is negative for
	\begin{equation}\label{eq:AQPQubit}
		\alpha^{\rm QP}_{d=2}=\frac{1}{1+|\lambda_{01}|}.
	\end{equation}
	Here, the superscript ${\rm QP}$ indicates that the corresponding bound was obtained via the negativity of the entanglement quasiprobability distribution.
	This result is identical with the condition for the negativity of the partial transposition [cf. Eq. \eqref{eq:ANPTD2Gauss}].
	As we consider the qubit case $(2\otimes2)$, all entanglement is NPT entanglement and it is not surprising that both methods yield the same result.
	The comparison also functions as a consistency check which shows that the entanglement quasiprobabilities provide correct results.
	
	\begin{figure}[t]
		\centering
		\includegraphics[width=\columnwidth]{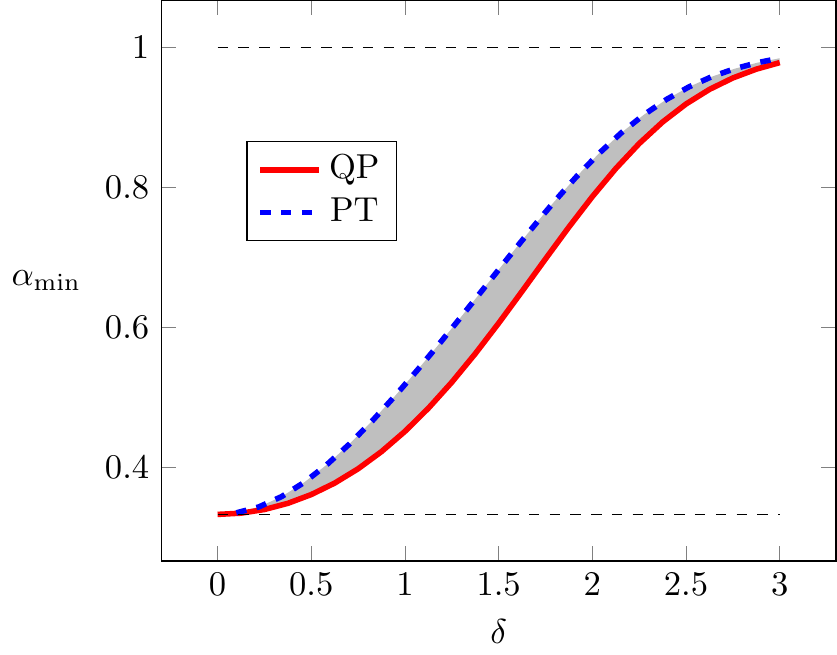}
		\caption{
		Verifying entanglement of qutrit Werner states undergoing Gaussian dephasing.
		Each curve shows the minimal value $\alpha_{\mathrm{min}}$ in dependence of the dephasing strength $\delta$ so that the corresponding criterion certifies entanglement for $\alpha>\alpha_{\mathrm{min}}$.
		The red solid line refers to the entanglement quasiprobabilities and the blue dashed line refers to partial transposition.
		The gray colored area between the solid and dashed line marks the entangled dephased qutrit Werner states with positive partial transposition.
		}
		\label{fig:QPPT}
	\end{figure}
	
	Let us now consider the qutrit case.
	From the qubit example we already see that one only has to find the minimal entry in the entanglement quasiprobability distribution and to show its negativity in order to certify entanglement.
	In the qutrit case the entanglement quasiprobability distribution has the following entries with $j\ne k$
	\begin{align}\label{eq:QPQutrit}
		\begin{aligned}
			p_{jk}^\parallel&=\frac{\mathcal{N}(\alpha)}{4}\left[1-\alpha\left(1+2|\lambda_{jk}|\right)\right],\\
			p_{jk}^\perp&=\frac{\mathcal{N}(\alpha)}{4}\left[1-\alpha\left(1-2|\lambda_{jk}|\right)\right],\\
			p_{jkm}&=\mathcal{N}(\alpha)\frac{1+\alpha}{8}.
		\end{aligned}
	\end{align}
	In the case of $|\lambda_{01}|\geq|\lambda_{02}|$, the smallest value is:
	\begin{equation}
		p_{01}^\parallel=\frac{\mathcal{N}(\alpha)}{4}\left[1-\alpha\left(1+2|\lambda_{01}|\right)\right].
	\end{equation}
	Note, that the absolute value of the dephasing coefficients $|\lambda_{jk}|$ depends on the difference $j-k$.
	Thus, we obtain the entanglement condition for the qutrit state
	\begin{equation}\label{eq:AQPQutrit}
		\alpha^{\rm QP}_{d=3}=\frac{1}{1+2|\lambda_{01}|}.
	\end{equation}
	Therefore, all dephased Werner states with $\alpha\geq\alpha^{\rm QP}_{d=3}$ cannot be given by a convex combination of separable states, and hence, are entangled.
	Comparing this result obtained from the entanglement quasiprobability with the NPT condition for the qutrit case, cf. Eq. \eqref{eq:ANPTGauss}, we find that both approaches deliver different results, i.e., $\alpha^{\rm QP}_{d=3}\neq\alpha^{\rm PT}_{d=3}$.
	Both results are plotted in Fig. \ref{fig:QPPT} for a Gaussian phase distribution.
	In particular, the gray shaded area in Fig. \ref{fig:QPPT} indicates the parameter region for which we identify bound entanglement, i.e., where we have negativities in the entanglement quasiprobabilities but positive partial transposition. 
	
	The set of bound entangled dephased Werner states with a Gaussian phase distribution is explicitly given by the parameters $\alpha$ which fulfill
	\begin{equation}\label{eq:BoundSet}
		\alpha^{\rm QP}_{d=3}(\delta)< \alpha\leq\alpha^{\rm PT}_{d=3}(\delta).
	\end{equation}
	Here, the dependence of the bounds on the dephasing strength $\delta$ is explicitly indicated [cf. also Eqs. \eqref{eq:ANPTGauss} and \eqref{eq:AQPQutrit}].
	Let us shortly consider the two limits of the this set of bound entangled states.
	In the case of $\delta=0$ all entanglement is verified by the partial transposition criterion, and hence, the set of bound entangled states is empty.
	For the other extreme, $\delta\to\infty$, all entanglement vanishes, which also yields an empty set of bound entangled states.
	However, for all other cases, $0<\delta<\infty$, the set in Eq. \eqref{eq:BoundSet} is not empty and there exist bound entangled states.
	The $\alpha$ interval is largest for $\delta=1.362$.

	This analysis demonstrates the strength of the entanglement quasiprobability method for entanglement verification when other methods fail.
	Specifically, we show that it can uncover bound entanglement in complex quantum states undergoing dephasing.
	The verification of entanglement in such disturbed systems is of great importance for quantum information applications under realistic conditions.


\section{Conclusion}\label{ch:Conclusion}

	In summary, we studied entanglement of dephased Werner states for which we could identify regions of bound entanglement.
	First, we derived the analytical expression of dephased Werner states for general dephasing and Gaussian dephasing, in particular.
	Second, we applied the partial transposition entanglement criterion to this family of noisy states, which allowed us to obtain the parameter regions for entangled states with negative partial transposition.
	Next we solved the separability eigenvalue problem of the dephased Werner states for qubit and qutrit states.
	Based on these solutions, we calculated the entanglement quasiprobability distributions for the qubit and qutrit case with general and especially Gaussian dephasing.
	
	By the negativities of the quasiprobabilities we were able to certify all entanglement of the studied states, as the negativities are a necessary and sufficient conditions for entanglement.
	Comparing the results in the qubit case, the partial transposition criterion and entanglement quasiprobabilities yield the same results.
	This serves as a consistency proof as the partial transposition criterion verifies all entanglement in the qubit scenario.
	For the qutrit case, we obtained quite different results with regard to the entanglement certification.
	Especially, we were able to verify bound entangled states by negative entanglement quasiprobability distributions for states with positive partial transposition.
	We derived an explicit parameter region of the bound entangled states.
	Our results indicate the strength of the method of entanglement quasiprobabilities for the verification of entanglement in complex and noisy states.
	Therefore, this method is a useful tool for verifying entanglement under realistic conditions.

\begin{acknowledgments}
	The authors are grateful to Stefan Gerke and Jan Sperling for enlightening discussions. 
	This paper has been supported by Deutsche Forschungsgemeinschaft through Grant No. VO 501/22-2.
\end{acknowledgments}

\appendix

\section{Solutions of the SEP}\label{app:Appendix}

Now we explain the derivation of the solutions of the separability eigenvalue problem of the dephased qubit and qutrit Werner states.
First, let us consider the Werner states in Eq. \eqref{eq:State}.
We formulate the corresponding separability eigenvalue problem in its second form, which yields
\begin{equation}\label{eq:AppSEPWerner}
	\hat\rho_W\ket{a,b}=\mathcal{N}(\alpha)(\ket{a,b}-\alpha\ket{b,a})=g\ket{a,b}+\ket{\chi},
\end{equation}
with the bi-orthogonal perturbation $\ket{\chi}$.
By expressing $\ket{b}$ with parallel and orthogonal projections with respect to $\ket{a}$ in the form
\begin{equation}
	\ket{b}=x\ket{a}+\sum_i x_i\ket{a_i^\perp},
\end{equation}
and $\ket{a}$ analogously with coefficients $y$ and $y_i$, the vector $\ket{b,a}$ can be written as
\begin{align}
	\begin{aligned}
		\ket{b,a}&=xy\ket{a,b}+\sum_{i,j} x_i y_j \ket{a_i^\perp,b_j^\perp}+\\
		&\quad x\sum_j y_j\ket{a,b_j^\perp}+y\sum_i x_i\ket{a_i^\perp,b}.
	\end{aligned}
\end{align}
Thus, with Eq. \eqref{eq:AppSEPWerner}, we have either $\ket{b,a}=\ket{a,b}$ and therefore $\ket{a}=\ket{b}$, or $\ket{b,a}=\ket{a^\perp,b^\perp}$ with $\ket{b}=\ket{a^\perp}$.
Therefore, the separability eigenvectors and -values of the Werner states are
\begin{align}
	\ket{a,b}&=\ket{a,a}\text{ with }g=\mathcal{N}(\alpha)(1-\alpha),\\
	\ket{a,b}&=\ket{a,a^\perp}\text{ with }g=\mathcal{N}(\alpha).
\end{align}
In the qubit case, a meaningful choice is
\begin{align}
	\ket{a,a}&=\ket{\mathfrak{s}_{01}^n,\mathfrak{s}_{01}^n},\label{eq:AppSEVQubit1}\\
	\ket{a,a^\perp}&=\ket{\mathfrak{s}_{01}^n,\mathfrak{s}_{01}^{n+2}},\ket{0,1},\ket{1,0},\label{eq:AppSEVQubit2}
\end{align}
with $n=0,1,2,3$.
These separability eigenvectors span the qubit Werner states.
Remember, that the dephasing only leads to a scaling of the off-diagonal elements while the algebraic structure is invariant.
Therefore, we insert these vectors into the separability eigenvalue problem of the dephased qubit Werner states.
We observe, that they are also the separability eigenvectors of the dephased states, if $\lambda_{01}\in{\mathbb R}$.
Hence, we use these separability eigenvectors for the calculation of an entanglement quasiprobability distribution for the dephased qubit Werner states as well.

For the qutrit cases, we use the same procedure.
As a set of separability eigenvectors for the qutrit Werner states we choose
\begin{align}
	\ket{a,a}&=\ket{\mathfrak{s}_{jk}^n,\mathfrak{s}_{jk}^n},\\
	\ket{a,a^\perp}&=\ket{\mathfrak{s}_{jk}^n,\mathfrak{s}_{jk}^{n+2}},\ket{\mathfrak{s}_{jk}^n,m},\ket{m,\mathfrak{s}_{jk}^n},
\end{align}
with $j,k=0,1,2$, $j\neq k$, $n=0,1,2,3$ and $m\neq j,k$.
These vectors span the qutrit Werner states and they are also separability eigenvectors of the dephased qutrit Werner states, if $\lambda_{ij}\in{\mathbb R}$. Hence we may use them for the calculation of the optimal entanglement quasiprobability distribution of the dephased qutrit Werner states.


\end{document}